# Selective diffraction with complex amplitude modulation by dielectric metasurfaces

Xu Song[1], Lingling Huang[1,2*], Chengchun Tang[3,4], Junjie Li[3,4], Xiaowei Li[5], Juan Liu[1], Yongtian Wang[1†], Thomas Zentgraf[2]

1. School of Optoelectronics, Beijing Institute of Technology, Beijing, 100081, China
2. Department of Physics, University of Paderborn, Warburger Straße 100, D-33098 Paderborn, Germany
3. Institute of Physics, The Chinese Academy of Sciences, Beijing, 100191, China
4. Beijing National Laboratory for Condensed Matter Physics, Beijing, 100191, China
5. Laser Micro/Nano-Fabrication Laboratory, School of Mechanical Engineering, Beijing Institute of Technology, Beijing 100081, China

**Abstract:**

Metasurfaces have attracted extensive interests due to their ability to locally manipulate optical parameters of light and easy integration to complex optical systems. Particularly, metasurfaces can provide a novel platform for splitting and diffracting light into several beams with desired profile, which is in contrast to traditional gratings. Here, we propose and experimentally demonstrate a novel method for generating independently selective diffraction orders. Our method is based on complex amplitude modulation with ultrathin dielectric metasurfaces. By tailoring the geometric parameters of silicon nanofin structures, we can spatially control the geometric and dynamic phase as well as the amplitude simultaneously. We compare the results with a metasurface that uses a phase-only modulation, to verify such selective diffraction can be solely efficiently achieved with complex amplitude modulation. Besides, the diffraction angles of each order have been measured, which are consistent with standard grating theory. Our developed method for achieving selective diffraction with metasurfaces has potential applications in beam shaping, parallel laser fabrication, and nanoscale optical detection.

---

\* Email: huanglingling@bit.edu.cn

† Email: wyt@bit.edu.cn

Metasurfaces have attracted great interests as novel kinds of flat artificial electromagnetic functional devices due to their unusual physical properties [1]. By patterning judiciously designed subwavelength nanoantennas or resonator along an optical surface, strong light interaction and local optical response modulation can be achieved. In contrast to conventional methods that utilize phase accumulations along optical paths, the abrupt phase shift can be realized within an extremely thin layer compared to the wavelength of light. Besides, the other local optical parameters, such as amplitude, polarization state, angular momentum or wave vector, can also be abruptly changed by metasurfaces, which make it possible to modulate a set of optical properties directly on the surface and in the scattered far-field. As a burgeoning research field, metasurfaces have shown great promise for novel design in a great number of device applications such as flat lenses [2-4], wave plates [5], beam deflectors [6, 7], switchable surface plasmon polariton couplers [8, 9], and high resolution three-dimensional holography [10-12].

Recently, a growing number of research on metasurfaces tend to replace the metallic (plasmonic) building blocks with all-dielectric resonators. Light scattering by high refractive index dielectric scatters has been shown to possess strong effective electric/magnetic dipole resonances, Kerker effects, or form birefringence [13]. One class of all-dielectric metasurfaces are reflectionless sheets, in which an overall pattern of the metasurfaces can be mimic Huygens' sources of waves, therefore also known as Huygens' metasurfaces [14]. Such kind of metasurfaces can be designed by balancing the electric and magnetic scattering responses to achieve impedance matching and hence minimizing the backward radiated power [15]. Another kind of dielectric metasurface is based on arrays of dielectric nanoposts or ridge waveguides, in which the required optical phase change is achieved over a subwavelength distance with high transmission efficiency [16, 17]. Using this concept, dielectric gradient phase metasurface optical elements based on the Pancharatnam-Berry (PB) phase have been proposed [18]. Such PB phase is a nontrivial phase modification result from beams traversing different polarization paths on the Poincaré sphere [19]. In fact, with dielectric metasurfaces, the wavefront modulation is indeed the combination of a PB phase and a dynamic phase from the nanostructures [20]. Benefited from extremely high transmission efficiency and easy integration into complex systems, all-dielectric metasurfaces can be used in various practical applications.

Traditionally, diffraction gratings are an effective way for beam control. By the adjustment of the grating period and duty cycle, diffraction orders and energy distributions can be tailored based on the Floquet boundary condition. Gratings have been applied in many fields, such as laser fabrication, optical communications, optical data storage, optical detection, and beam shaping [21-23]. As an improvement to conventional gratings, Dammann gratings, by the phase optimization of each individual pixels within a supercell (which is the smallest periodic cell of the grating), are capable of generating a uniform or tailored intensity distribution in a number of one-dimensional (1D) and two-dimensional (2D) diffraction orders [24]. However, the energy distribution of the diffraction orders of Dammann gratings relies on the proper setting of the evaluation functions. Several methods such as simulated annealing algorithm and genetic algorithm for the extensions of the Dammann optimization have been proposed for more flexibility in the intensity distribution of diffraction orders [25]. Nevertheless, it is still difficult completely suppressing the intensities of all the unselected diffraction orders. As a result, the diffraction orders are not independently controllable. In addition, the diffraction efficiency is relatively low due to the phase-only modulation and the size of traditional Dammann device is still relatively large for an integrated optical system. The possible solution to achieve selective diffraction is to replace the binary phase modulation with complex amplitude and phase modulation. Therefore, it is desirable to combine the concept of Dammann optimization algorithm with the flexibility of metasurfaces for the achievement. Actually, several metasurfaces based on Dammann principle have been utilized for generation of optical vortex arrays [26, 27] and beam arrays [20, 28].

In this letter, we propose and experimentally demonstrate a novel approach for the generation of two-dimensional selective diffraction orders based on tailored dielectric metasurfaces with complex amplitude modulation using the PB phase principle. By positioning the azimuthal angles of dielectric nanofins and simultaneously tuning the geometry parameters (length and width) of each nanofin, the generated phase and amplitude can be directly modulated. Such dielectric nanofins can generate different dynamic phase delays for different geometry parameters even with the same orientation angle. Therefore, each nanofin has its own initial transmission phase change, which will be combined with the PB phase that results from the azimuthal angle to achieve the final desired local phase modulation. For circularly polarized (CP) light, such metasurface can generate arbitrary 2D diffraction orders based on a complex amplitude modulation. Figure 1 schematically illustrates the generation and reconstruction procedure of 2D selective diffraction orders based on our proposed dielectric metasurface. Silicon nanofins are arranged on a glass substrate with the same lattice constants in both directions. For experimental validation of the proposed concept, we designed and fabricated a sample for which the selected diffraction orders form a pattern with the word "META" while all other diffraction orders are simultaneously suppressed. In our experiment, we analyzed the distribution uniformity and angles of the diffraction orders and compared them with a phase-only design. We found that the simulation results and the experimental far-field images are well consistent with each other, confirming the possibility to achieve selective diffraction with dielectric metasurfaces.

The basic unit-cell of the metasurface is composed of a Si dielectric nanofin on top of a fused quartz substrate (Figure 2a). The lattice constant of the rectangular nanofins is 300 nm for both *x*-axis and *y*-axis, which is much less than the desired incident wavelength of 785 nm. To acquire the complex amplitude modulation between the two orthogonal circular polarization states, we carry out a 2D parameter optimization using a rigorous coupled wave analysis method to optimize the geometry parameters of the nanofins. The height is set to 380 nm, while the length and width are swept in the range of 80 nm to 200 nm and 40 nm to 150 nm, respectively. The refractive index of the fabricated Si was measured by ellipsometry, which is $n_\lambda$=3.8502+0.0109*i* with respect to the wavelength $\lambda_{in}$ =785 nm. The calculated transmission amplitude and phase delay for the conversion transmission coefficients $T_{lr}$ (for left to right circular polarization) are shown in Figure 2b and Figure 2c. When the length *L* of the nanofin is 190 nm and the width *w* is 100 nm, the amplitude of $T_{lr}$ can reach up to 0.9589. If all nanofins are designed with this geometry parameter, the incident circularly polarized light beam will be almost fully transformed into the opposite helicity, therefore such a metasurface will work as a half-wave plate [18]. Besides, the amplitude may take any value from 0 to the maximum by choosing suitable geometry parameters. Here we choose five distinct amplitude values with 0.2, 0.4, 0.6, 0.8 and 0.9589 as shown in Figure 2d. Each of them corresponds to a pair of characteristic length and width for the nanofin, which are marked with black circles in Figure 2b. Figure 2c shows the corresponding phase delay for $T_{lr}$ of the selected nanofins. Notice that for the selected geometry parameters, the extra transmission phase shifts of the nanofins differ from each other. Apart from the dynamic phase delay due to the accumulation within the dielectric nanofin, there exists a geometric PB phase Φ related to the azimuthal angle of the nanofin with respect to the *x*-axis. Such geometry phase corresponds to half of the solid angle encompassed by the geodesic path of the final and initial polarization state on the Poincaré sphere. Hence, the phase shift Φ of the orthogonal handedness CP light is twice of the orientation angle $\theta$ of the nanofin, which can be express as $\Phi = 2\sigma\theta$, whereas $\sigma$ represents the incident helicity of the light. Since each of the distinct rectangular nanofins corresponds to different dynamic phase shifts, an extra angle rotation must be induced into distinct selected nanofins for compensation. The finally desired phase $\alpha_d$ of each pixel is the superposition of the PB phase and compensation phase $\alpha_c$ which is equal to the dynamic transmission phase shift. Accordingly, the orientation angle $\theta$ of the nanofin can be designed as follows:

$$\theta = (\alpha_d - \alpha_c)/2 \qquad (1)$$

When light normally illuminates upon the metasurface with the desired wavelength, the diffraction orders and diffraction angles are determined only by the supercell of the metasurface, which is arranged periodically in two dimensions utilizing the Dammann grating concept. Here we choose a supercell composed of 60×60 nanofins, with a lattice constant of 300 nm, resulting in a supercell area of 18 μm×18 μm. The whole metasurface contain 13×13 supercells. The diffraction orders will be controlled by the design of the nanofin arrays within the supercell for achieving the complex amplitude modulation. The desired diffraction orders can be expressed as the superposition of Fourier expansion according to the formula as follows,

$$U(x,y) = \sum_{p,q} \exp(i2\pi(px/d_x + qy/d_y)) \quad (2)$$

where $p$ and $q$ are the indices of the diffraction orders, $d_x$ and $d_y$ are the size of the supercell in each direction. To form the pattern of the word "META" with the different diffraction orders, $p$ and $q$ are selected in the range of -5 to 5 and -7 to 7, respectively (see Figure 1). Note that for all the selected diffraction orders, the amplitude has the maximum U=1 at coordinate $(x,y) = (nd_x/p, md_y/q)$, whereas $n$ and $m$ are arbitrary integers. If all the maximum values are superimposed in the same unit-cell of each supercell, it will be difficult to normalize the amplitudes into limited stages since the peak value is much larger than all the other values. Here we made a phase shift for different diffraction orders so that the maximum values are distributed to different pixels in the supercell, which will result in an amplitude as follows:

$$U(x,y) = \sum_{p,q} \exp(i2\pi(p(x+ql_c)/d_x + q(y+pl_c)/d_y)) \quad (3)$$

where $l_c$ represents the lattice constant of the unit-cell. Next, each nanofin is designed and fabricated by mapping to the 5-stage amplitude and continuous phase modulation (sample B). For comparison, we also fabricated a metasurface with phase-only modulation for which all the nanofins are chosen with the equal amplitude of 0.9589 (sample A). The theoretically calculated phase for both samples are shown in Figure 3a and Figure 3b. The samples were fabricated on fused quartz plate using electron beam lithography, followed by a plasma etching. The scanning electron microscopy (SEM) image of the fabricated phase-only sample A and complex amplitude sample B are shown in Figure 3c and Figure 3d, respectively. The SEM images clearly show that different geometric parameters for the nanofins of sample B.

The fabricated samples are experimentally characterized for their performance with the experimental setup shown in Figure 3e. The incident light with λ=785 nm passes through a linear polarizer and a quarter-wave plate to form the desired CP light. For the characterization, two sets of experimental configurations are used for measuring and imaging the diffraction orders. On one hand, for measuring the diffraction angles of the different orders, a rotation stage with a quarter-wave plate and a linear polarizer are set behind the metasurface sample. By using a 2D translation stage in *xy*-plane, the precise location and intensity of each diffraction order were measured. On the other hand, for the purpose of completely imaging all the selective diffraction orders together, all the beams are collected by a microscope objective and imaged to a CCD camera. The numerical aperture of the objective is NA=0.45, which allows imaging the diffraction angles up to ±26.74° with respect to the *z*-axis. The maximum diffraction orders from such metasurface can be determined according to the grating theory [29],

$$\left|\mathbf{k}_\parallel^{in} + pK\hat{x} + qK\hat{y}\right| = n_0 k_0 \quad (4)$$

where $\mathbf{k}_\parallel^{in}$ is the projection of the incident wave vector in the *xy*-plane. $\hat{x}$ and $\hat{y}$ are unit vectors in the two supercell directions, $k_0 = 2\pi/\lambda$ is the vacuum wave number, and $n_0$ is the refractive index of air. Since the supercell periodicity is the same in two dimensions $d_x = d_y = d$, the grating vector is given by $K = 2\pi/d$. For normal illumination perpendicular to the metasurface, the polar propagation angle $\psi_{pq}$ of the emerging beams with respect to the optical axis (*z*-axis) can be expressed as

$$\sin^2 \psi_{pq} = (p^2 + q^2)\frac{\lambda^2}{d^2} \quad (5)$$

For the pattern of the word "META", the maximum diffraction orders are (±5, ±7). With the designed working wavelength of 785 nm and a periodic supercell of 18 μm in both directions, the largest angle of the designed diffraction orders with respect to the *z*-axis is 22.03°.

The theoretical simulation results for the selected diffraction orders reconstructed from the metasurfaces for both the original complex amplitude and the phase-only patterns are shown in Figure 4a and Figure 4b, respectively. Obviously, unwanted diffraction orders emerge if the amplitude information is eliminated for the phase-only pattern. However, when the original amplitude values are normalized and replaced by the 5-stage values while keeping the continuous phase modulation, all the selected orders can be obtained with very good consistency compared to the results for the ideal complex amplitude, as shown in Figure 4c. In addition, we performed for the 5-stage complex amplitude modulation a far-field simulation based on finite-difference time-domain method for analyzing the polarization conversion in full 3D space. In order to reduce the computation load, only one period of the supercell is simulated. As shown in Figure 4d, each of the selected diffraction orders converts to the orthogonal polarization state of the incident light while the intensity of other locations is very weak with disordered polarization states. For clarity, irrelevant information with little energy distribution has been omitted in this figure.

The experimental results are shown in Figure 4e to 4h. Actually, for the phase-only design (sample A), the modulations only depend on the azimuthal orientation controlled by PB phase, hence it can be considered as a broadband device. The results of sample A for two different wavelengths of 633 nm and 785 nm are shown in Figure 4e and 4f, respectively. The unwanted diffraction orders with much lower energy emerge as expected, while all the selected diffraction orders forming the words "META" are still easy to be identified at both wavelengths. Because all the nanofins are chosen with a high transmission, the desired diffraction orders have the intensity much higher than the others. However, the designed sample B for complex amplitude modulation is effective only for the desired wavelength of 785 nm because the total phase is applicable only for a single wavelength. Therefore, for sample B, the designed diffraction orders cannot be distinguished from the other orders at the wavelength of 633 nm (Figure 4g). At the design wavelength of 785 nm the result shows all the selected diffraction orders very well, compared with both the results of sample A in Figure 4f and 4e. There are still a small number of unwanted diffraction orders that can be observed with weak intensities. These deviations might be the result of fabrication errors of the nanofins and slight differences of material parameters. For sample B, the measured diffraction efficiency of the opposite helicity with respect to the incident light at 785 nm is 28.21%, which is close to 31.02% from our simulation result. After subtracting the energy of the zero-order diffraction, the average efficiency of the 65 desired selective orders with respect to the orthogonal CP light at 785 nm is about 0.31% for each order. For the wavelength of $\lambda=633$ nm several undesired diffraction orders appear. Hence, the determination of the total efficiency for this wavelength is meaningless. We note that the zero-order light in the experiments results from a small polarization change of the used quartz substrate, which is slightly anisotropic.

The diffraction angles are also measured in the experiment. Figure 5a shows a superposition of the diffraction orders in the far field when illuminating metasurface sample A with both wavelengths (633 nm and 785 nm) simultaneously. As expected by equation 5, each diffraction order at 785 nm has a slightly larger diffraction angle compared with the one at 633 nm. Figure 5b shows the selected diffraction orders with schematically on a spherical surface with the chosen notation for the angles. Note that the diffraction angles measured of sample A and B are the same due to the identical supercell size. In our design, the orders are concentrated in a small cone area with respect to the *z*-axis, which makes them easier to be collected by objective. Actually, the diffraction angles can be designed in a much larger angular range by setting the proper lattice constant and pixel numbers of the supercell of the metasurface. In the following we analyze the propagation direction of the diffraction orders, which can be denoted by a polar angle $\psi$ and an azimuthal angle $\zeta$. They correspond to the angles of the wave vector with respect to the z-axis and the projection of wave vector on the xy-plane with respect to the x-axis. For normally incident light, the incident wave vector is $k_0 = \alpha_0 \hat{x} + \beta_0 \hat{y} - \gamma_{00} \hat{z}$ [30], where $\alpha_0 = k_0 \sin\psi_0 \cos\zeta_0$, $\beta_0 = k_0 \sin\psi_0 \sin\zeta_0$ with $\psi_0 = 0, \zeta_0 = 0$. For the diffraction order (p,q), we have $\alpha_p = \alpha_0 + 2\pi p/d_x = 2\pi p/d$ and $\beta_q = \beta_0 + 2\pi q/d_y = 2\pi q/d$. And $\gamma_{mn}$ can be obtained by $\alpha_m^2 + \beta_n^2 + \gamma_{mn}^2 = k_0^2 n_0^2$. $\psi$ can be calculated from equation 5 and $\zeta$ can be expressed as [31]

$$\zeta_{pq} = \arctan \frac{q d_x}{p d_y} \qquad (6)$$

These diffraction parameters were calculated from experimental data, which gives the precise location of each diffraction order with respect to the coordinate axis,

$$\begin{cases} \psi = \arccos(l/\sqrt{a^2 + b^2 + l^2}) \\ \zeta = \arccos(a/\sqrt{a^2 + b^2 + l^2}) \end{cases} \qquad (7)$$

Here a and b are the locations of the measured points with respect to the zero-order along the optical axis in x-axis and y-axis, respectively. l is the distance between the measured points and the surface of the sample along the z-axis. The diffraction angles of $\psi$ and $\zeta$ by theoretical calculations and experimental measurements are shown in Figure 5c, respectively. The angles with respect to different diffraction orders in two dimensions are plotted with different (p, q) orders in distinct colors. The theoretical and experimental data are represented as boxes and asterisks, respectively. Clearly, the measured values and theoretical calculation are consistent with each other for all diffraction orders. Therefore, we can achieve 2D beam distributions in arbitrary directions accordingly.

We note that a number of feasible metasurfaces have been proposed to modulate complex amplitude for transmitted light, such as C-shape nanoantennas [32] or double layers composed of L-shape slot and concentric loop resonators [33]. However, most of them work in Terahertz or microwave range. With the simplification of such tailored nanofins, it is easy to independently modulate both the phase and amplitude in the shorter visible and near-infrared range. Meanwhile, the combination of dynamic phase and geometric phase for the phase design principle is crucial to the successful implementation. Such metasurface can achieve selective diffraction orders by breaking the structural symmetry compared to traditional gratings, providing more design freedoms for various parallel beam controlling applications, with predefined directions and polarizations, which is not easily achievable with other methods.

To summarize, a novel design strategy for generating selective diffraction orders based on dielectric metasurface with complex amplitude modulation of the incident light is proposed and experimentally demonstrated. For each of the subwavelength nanofin, the amplitude can be manipulated by modulating the geometric parameters in several normalization stages and the desired phases are continuously controlled by the orientation angles of the nanofins. The two-dimensional selective diffraction orders were generated based on the superposition of Fourier expansion, which can strongly suppress the unwanted diffraction orders. Both simulation and experimental results are in agreement with each other, which verifies our complex amplitude design at the desired wavelength. The diffraction angles of the different diffraction orders were experimentally measured which were in accordance with theoretical calculations. The proposed method can be used for potential applications in beam shaping, parallel laser fabrication, laser cooling, and nanoscale optical detections and so on.

## Fabrication methods

Plasmonic meta-surfaces were fabricated on a fused quartz plate using electron beam lithography, followed by a plasma etching. Firstly, a 380 nm thick silicon film was deposited by plasma enhanced chemical vapor deposition (PECVD) method, then a PMMA film with a thickness of 200 nm was spinning coated at 4000 RPM for 60 s and was covered by 35nm PEDOT:PSS film as conducting layer. The desired structure was directly written by JEOL 6300FS EBL at 1150 μC/cm$^2$ with an accelerating voltage of 100 kV. After the exposure process, the PEDOT:PSS layer was washed away with pure water and the resist was developed in 1:3 MIBK:IPA solution for 40 s. After development, the sample was washed with IPA and baked at 110℃ for 120 s before coating 80 nm Cr by electron beam evaporation deposition (EBD) method. For the purpose of lift-off, the sample was immersed in hot acetone of 75 ℃, followed by ultrasonic cleaning with acetone ethanol and pure water for 20 s respectively. Finally, the desired structure was transferred from Cr to silicon by inductively coupled plasma etching (ICP) method with HBr as a reactive gas and the remaining Cr was removed by wet erosion method.

## Acknowledgments

The authors acknowledge the funding provided by the National Natural Science Foundation of China (No. 61775019, 61505007) program and Beijing Municipal Natural Science Foundation (No. 4172057). L.H. acknowledge the support from Beijing Nova Program (No. Z171100001117047) and Young Elite Scientists Sponsorship Program by CAST (No. 2016QNRC001). This project has received funding from the European Research Council (ERC) under the European Union's Horizon 2020 research and innovation programme (ERC grant agreement No. 724306).

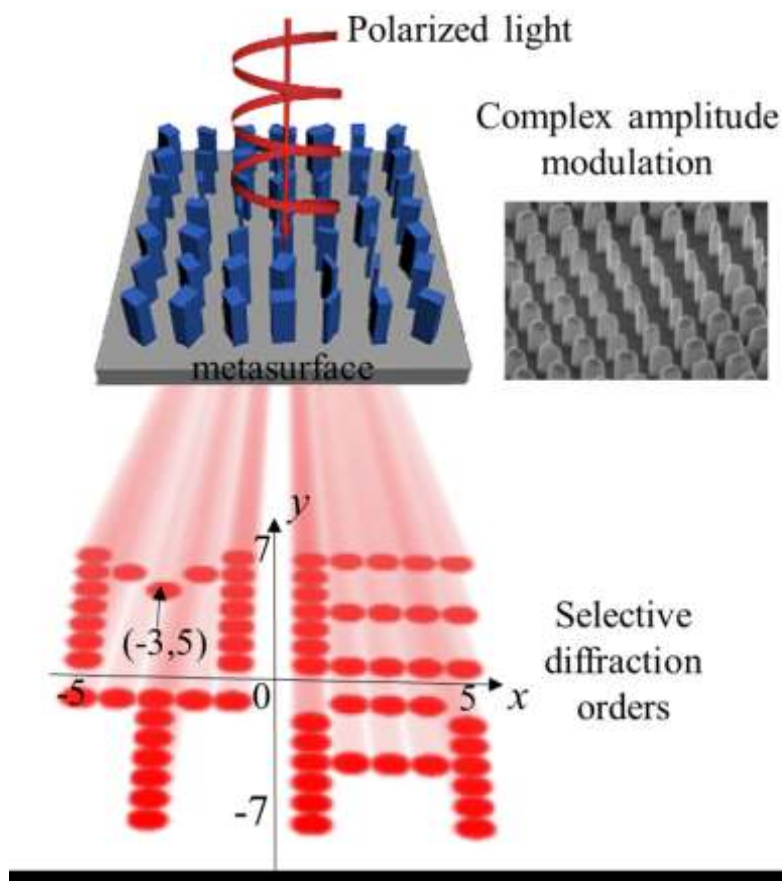

Figure 1. Schematic illustration of the complex amplitude modulation for selective diffraction with a dielectric metasurface. The diffractions orders are chosen to show a pattern of the word "META". The SEM image shows a typical example of a dielectric metasurface made of silicon nanofins.

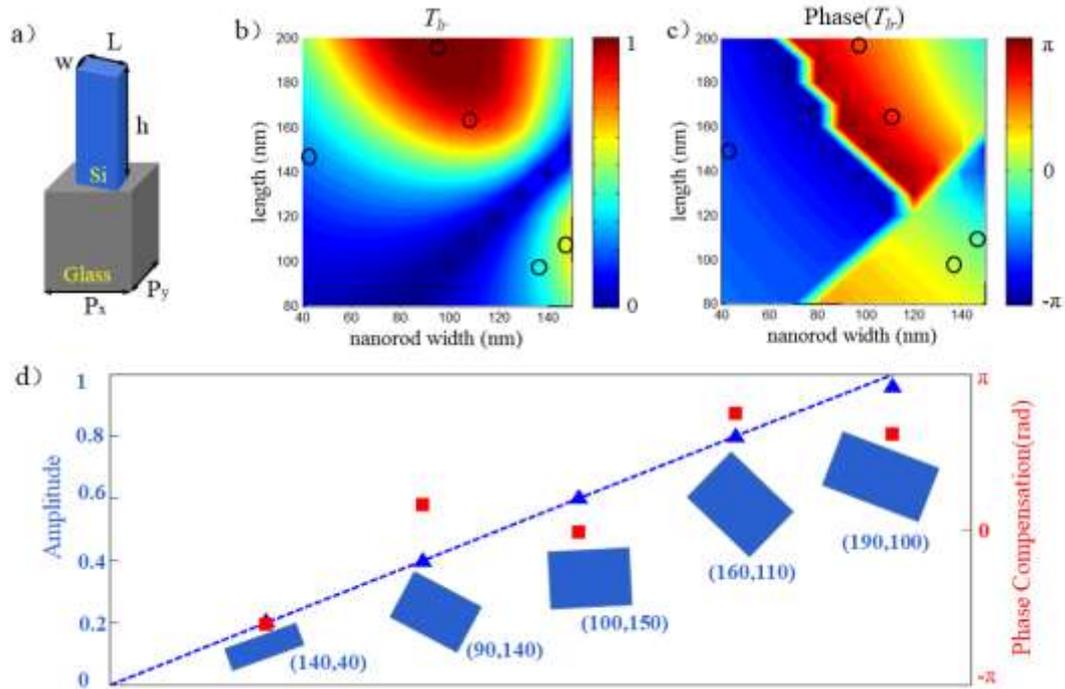

Figure 2. (a) Schematic illustration of a single Si nanofin unit-cell above glass substrate. The geometry parameters and orientation angle of the nanofin can be tailored for complex amplitude modulation. (b) The amplitudes and (c) phase delays for the cross-polarization transmission coefficients $T_{lr}$ with respect to different geometry parameters of the nanofin. The values are obtained by RCWA simulation for a wavelength of 785 nm. The selected geometry parameters of the nanofins are marked with black circles. (d) Transmission amplitude and phase shifts for the selected nanofins at the design wavelength of 785 nm. The rotated patterns represent the nanofins for the corresponding five stage of complex amplitude modulation. The values in the parenthesis correspond to the length and width of the nanofins, whereas the orientation angles determine the phase compensation by applying the PB phase principle for the orthogonal CP light.

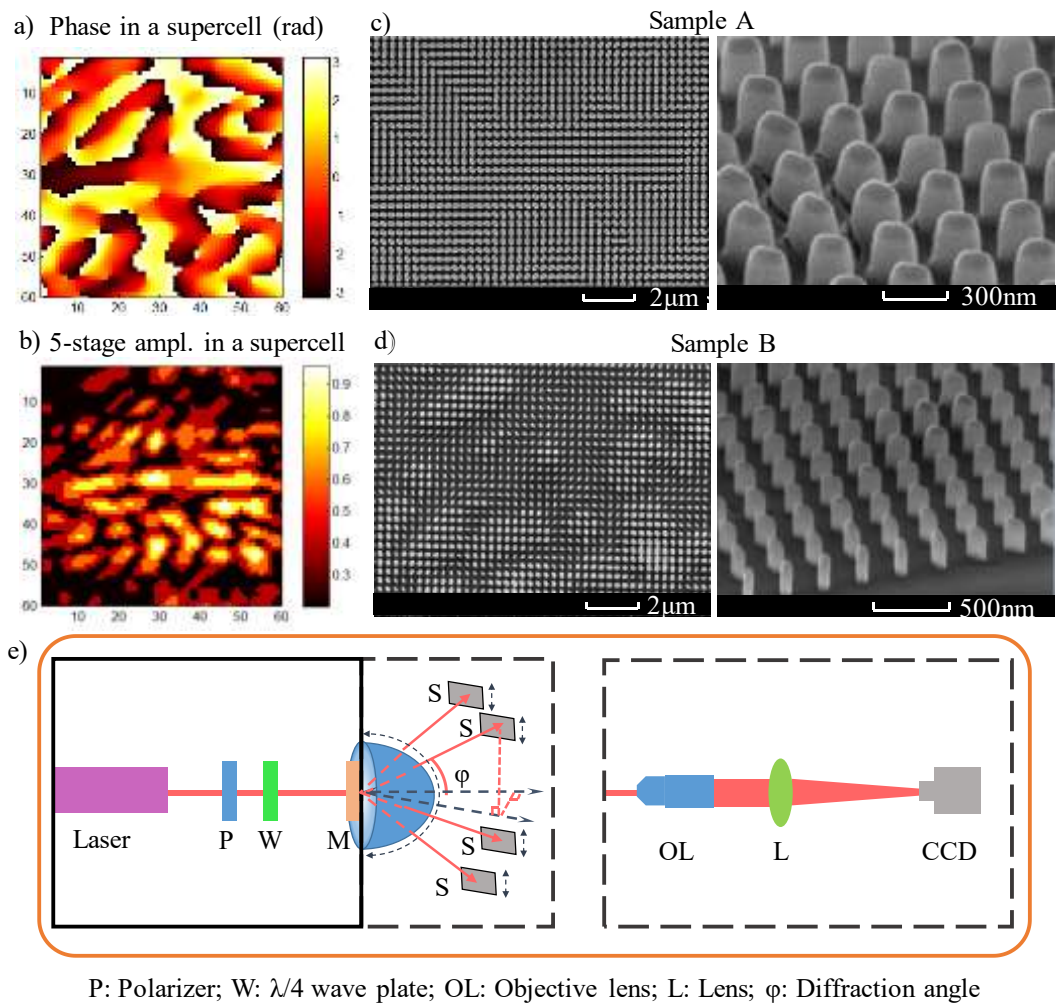

P: Polarizer; W: λ/4 wave plate; OL: Objective lens; L: Lens; φ: Diffraction angle

Figure 3. (a) Theoretically calculated phase distribution of both samples A and B. The *x*- and *y*-axis correspond to the number of the unit-cell in the respective direction. (b) Calculated 5-stage amplitude modulation with normalization of sample B. (c) SEM images of the fabricated metasurface sample A in top view and side view. (d) SEM images of fabricated metasurface sample B. (e) Experimental setup for measuring the angles of the diffraction orders and imaging of the selected orders.

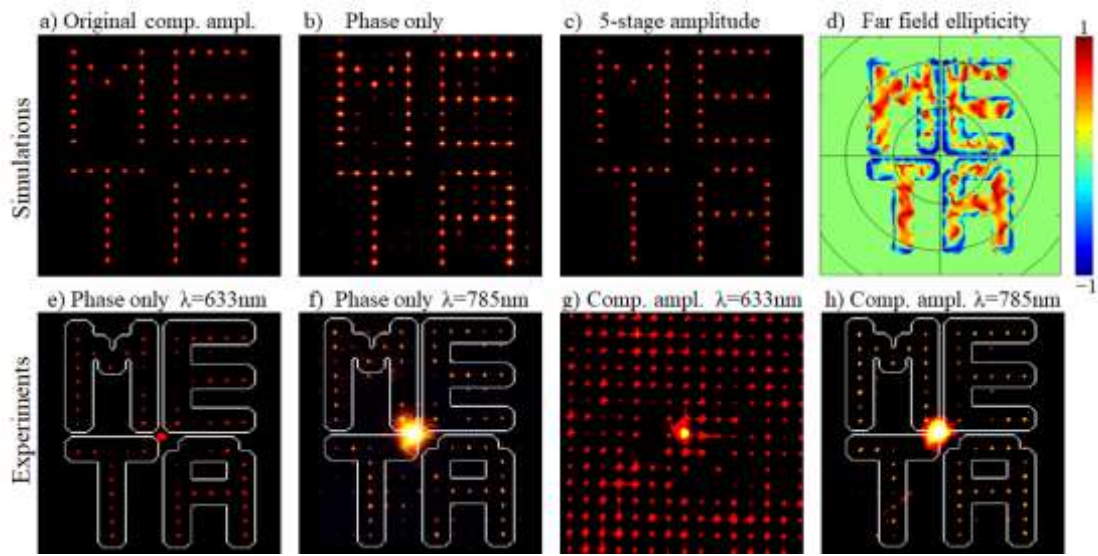

Figure 4. (a) Numerical reconstruction of the diffraction orders from the original complex amplitude target pattern. (b) Simulated diffraction results with the phase-only modulation. Unwated diffraction orders appear in the image. (c) Simulated diffraction result with a continuous phase and normalized 5-stage amplitude modulation. (d) FDTD simulation result with one supercell of the metasurface for far-field ellipticity. (e,f) Experimental results of sample A with phase-only modulation at 633 nm and 785nm. (g,h) Experimental results of sample B with complex amplitude modulation at 633nm and 785nm.

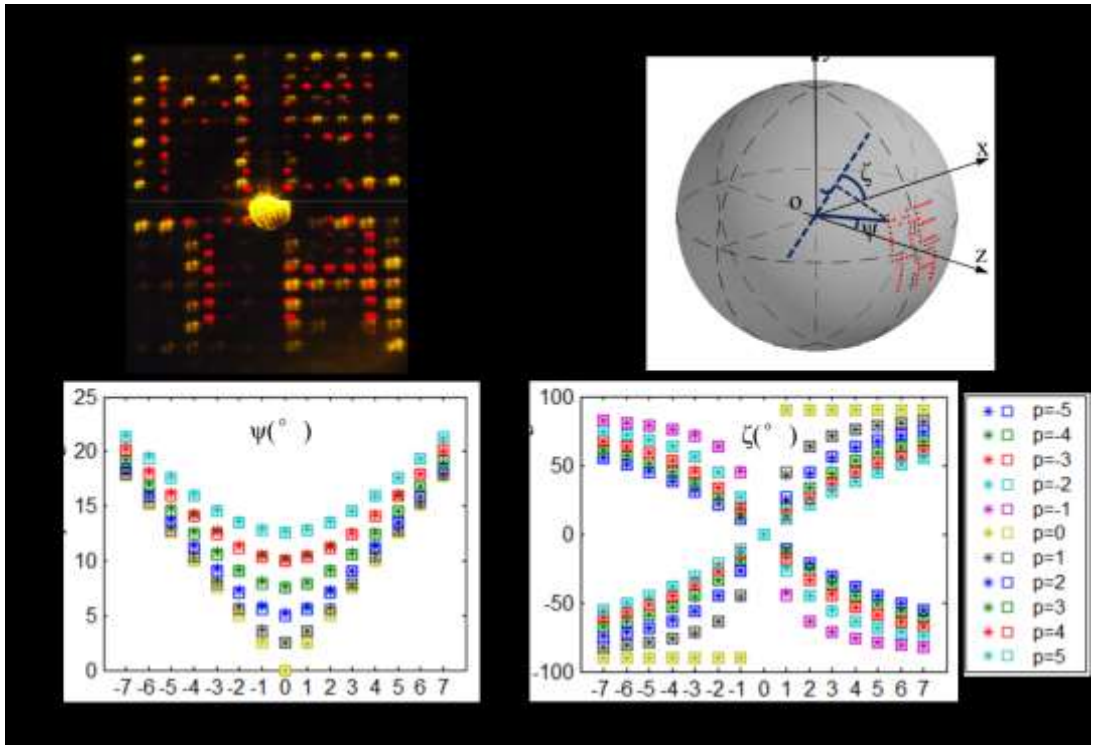

Figure 5. (a) Experimental results for sample A with phase-only modulation as a superposition of the two wavelengths of 633 nm (red dots) and 785 nm (golden dots). (b) Schematic view of the angles for diffraction orders in spherical coordinates. (c) Results for the angle of the wave vector $\psi$ with respect to the $z$-axis (left) and the angle of the projection of wave vector $\zeta$ on $xy$-plane with respect to the $x$-axis (right) for all diffraction orders at the wavelength of 785 nm. The theoretical and experimental values marked by the boxes and the asterisks, respectively.